\documentclass[lefttitle,3p]{elsarticle}
\usepackage[utf8]{inputenc}
\usepackage{textgreek}
\usepackage{siunitx}
\usepackage{caption}
\usepackage{subcaption}
\usepackage{amsmath}
\usepackage{amssymb}
\usepackage{amsfonts}
\usepackage{amsthm}
\usepackage{natbib}
\usepackage{import}
\usepackage{xifthen}
\usepackage{pdfpages}
\usepackage{transparent}

\journal{Nuclear Instruments and Methods in Physics Research}

\graphicspath{{fig/}}

\newcommand\NOvA{NO\textnu A}

\newcommand\chips{\textsc{Chips}}

\newcommand\chipsfive{\textsc{Chips-5}}

\begin{document}

\begin{frontmatter}
    %%%%%%%%%%%%%%%%%%%%%%%%%%%%%%%%%%%%%%%%%%%%%%%%%%%%%%%%%%%%%%%%%%%%
% Paper title
%%%%%%%%%%%%%%%%%%%%%%%%%%%%%%%%%%%%%%%%%%%%%%%%%%%%%%%%%%%%%%%%%%%%
\title{Low-latency NuMI Trigger for the \chipsfive{} Neutrino Detector}

%%%%%%%%%%%%%%%%%%%%%%%%%%%%%%%%%%%%%%%%%%%%%%%%%%%%%%%%%%%%%%%%%%%%
% Authors
%%%%%%%%%%%%%%%%%%%%%%%%%%%%%%%%%%%%%%%%%%%%%%%%%%%%%%%%%%%%%%%%%%%%
\author[1]{Simeon~Bash}

\author[1]{John~Cesar}

\author[3]{Greg~Deuerling}

\author[1]{Thomas~Dodwell}

\author[1]{Stefano Germani}

\author[1,2]{Petr~Mánek\corref{cor1}}
\ead{petr.manek.19@ucl.ac.uk}

\author[3]{Evan~Niner}

\author[3]{Andrew~Norman}

\author[1]{Jennifer~Thomas}

\author[1]{Josh~Tingey}

\author[3]{Neil~Wilcer}

%%%%%%%%%%%%%%%%%%%%%%%%%%%%%%%%%%%%%%%%%%%%%%%%%%%%%%%%%%%%%%%%%%%%
% Affiliations
%%%%%%%%%%%%%%%%%%%%%%%%%%%%%%%%%%%%%%%%%%%%%%%%%%%%%%%%%%%%%%%%%%%%
\address[1]{Department of Physics and Astronomy, University College London, Gower Street, London, WC1E 6BT, United Kingdom}
\address[2]{Institute of Experimental and Applied Physics, Czech Technical University, Husova 240/5, Prague, 110 00, Czech Republic}

\address[3]{Scientific Computing Division, Fermi National Accelerator Laboratory, Illinois, United States}

\cortext[cor1]{Corresponding author}

% TODO: elsarticle template recommends using \afilliation but it does not work, figure out why

%%%%%%%%%%%%%%%%%%%%%%%%%%%%%%%%%%%%%%%%%%%%%%%%%%%%%%%%%%%%%%%%%%%%
% Abstract
%%%%%%%%%%%%%%%%%%%%%%%%%%%%%%%%%%%%%%%%%%%%%%%%%%%%%%%%%%%%%%%%%%%%
\begin{abstract}
    The \chips{} R\&D project aims to develop affordable large-scale water Cherenkov neutrino detectors for underwater deployment.
    In~2019, a \SI{5}{\kilo\tonne} prototype
    detector \chipsfive{} was deployed in northern Minnesota to potentially study neutrinos
    generated by the NuMI beam.
    This paper presents the dedicated low-latency triggering system for \chipsfive{} that delivers
    notifications of neutrino spills from the Fermilab accelerator complex to the detector with sub-nanosecond precision. Building on existing \NOvA{} infrastructure, the time distribution system achieves this using only open-source software and conventional computing and network elements. In a time-of-flight study, the system 
    reliably provided advance notifications~$610\pm\SI{330}{\milli\second}$ prior to neutrino spills at 96\% efficiency. This permits advanced analysis in real-time as well as hardware-assisted triggering that saves data bandwidth and reduces DAQ computing load outside time windows of interest.
\end{abstract}

%%%%%%%%%%%%%%%%%%%%%%%%%%%%%%%%%%%%%%%%%%%%%%%%%%%%%%%%%%%%%%%%%%%%
% Keywords
%%%%%%%%%%%%%%%%%%%%%%%%%%%%%%%%%%%%%%%%%%%%%%%%%%%%%%%%%%%%%%%%%%%%

\end{frontmatter}

\section{Introduction}

\begin{figure}
    \centering
    \includegraphics[width=0.35\linewidth]{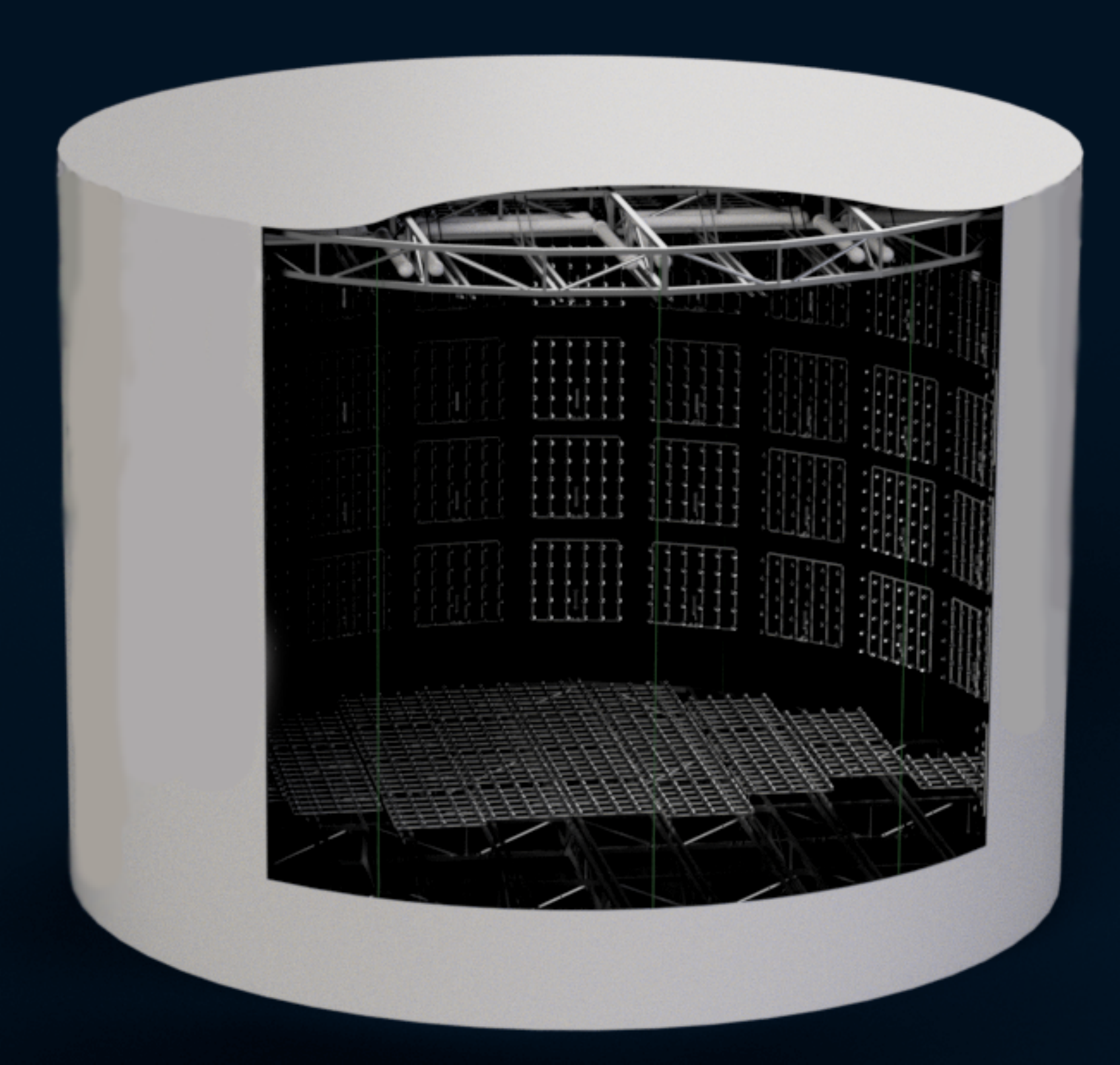}
    \caption{Artist's rendering of the \chipsfive{} detector.}
    \label{fig:detector-render}
\end{figure}

The purpose of the \chips{} project is to develop affordable water Cherenkov neutrino detectors suitable for large-scale distributed deployments. Aiming to produce results of comparable quality to other current water Cherenkov experiments, \chips{} seeks to significantly reduce the cost per unit mass by utilising off-the-shelf components, readily available materials, and by designing modular parts that can be rapidly scaled and easily serviced. In 2019, \chipsfive{}, the latest-generation \chips{} detector prototype, was deployed in northern Minnesota, USA. Comprising a \SI{12}{\meter}~tall cylindrical volume of \SI{12.5}{\meter}~radius (illustrated in Figure~\ref{fig:detector-render}) the detector can provide up to \SI{1924}{\square\meter} available surface area, offering the first practical insights into construction, deployment and operation of \chips{} devices at such scale.

The site in northern Minnesota lies \SI{7}{\milli\radian} off the axis of the NuMI beam~\cite{adamson2016}, a well-understood source of muon neutrinos that originates at Fermilab, Illinois.  Under normal operation, NuMI delivers neutrinos in periodic spills separated by time intervals filled with cosmic background that would dominate the data stream if recorded without interruption. This motivates the introduction of the triggering system described here. Other detectors that have observed NuMI neutrinos such as MINOS~\cite{michael2003minos} or \NOvA{}~\cite{ayres2005nova} have applied the timing signals after the fact~\cite{romisch2012synchronization,norman2012nova}, during the offline processing. Apart from using the accurate time-stamping capability developed for \NOvA{}, the \chipsfive{} trigger uses only consumer electronics and conventional network elements that are affordable and widely available. Since triggering is performed in real-time, fractions of a second prior to neutrino arrival at the detector site, emphasis is put on the overall latency of the system.

The rest of the paper is organised as follows. Firstly, section~\ref{sec:nova-tdu} provides background on accelerator signals that are used as a basis for the trigger, and describes the facilities required for their decoding. Secondly, section~\ref{sec:signal-delivery-chain} focuses on the timing distribution system that delivers notifications from the accelerator complex over the baseline distance to the \chipsfive{} detector site. Thirdly, section~\ref{sec:onsite-triggering} gives details on different triggering methods implemented on-site. Section~\ref{sec:benchmark} presents experimental measurements that demonstrate the viability of the new system and discusses the results.

\section{Implementation}

The trigger system described here is implemented as a chain with several stages that operate simultaneously together with the Data Aquisition (DAQ) at the detector site. They are described in order here.

%%%%%%%%%%%%%%%%%%%%%%%%%%%%%%%%%%%%%%%%%%%%%%%%%%%%%%%%%%%%%%%%%%%
\subsection{Neutrino Spill Detection}%
\label{sec:nova-tdu}

NuMI neutrino spills can be detected by observing events of interest occurring in components of the accelerator. At Fermilab, these events are tracked by two precise timing systems:
\begin{enumerate}
    \item The Beam-synchronous Clock System (BSYNC).
    \item The Tevatron Clock (TCLK)~\cite{beechy1986time}.
\end{enumerate}

\begin{figure}
    \centering
    \includegraphics[width=0.8\linewidth]{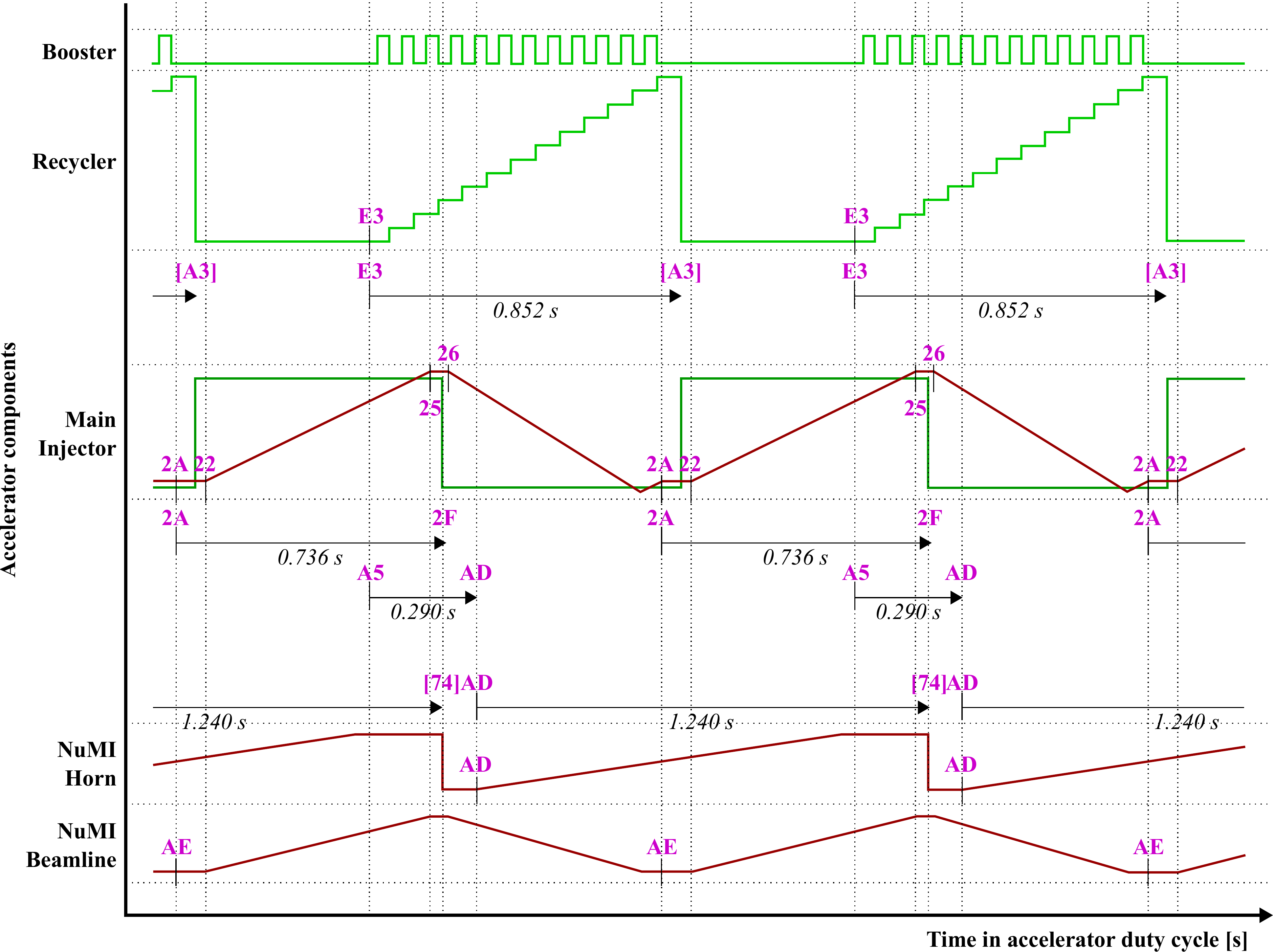}
    \caption{Operation of selected parts of the accelerator duty cycle relevant to the NuMI beam plotted as a function of time. Different series illustrate the current delivered to control magnets of various components (not to scale). Accelerator signals are labeled in bold magenta (e.g.~AE, 2A).~\cite{PhilAdamsonDiagram}}
    \label{fig:numi-time-signals}
\end{figure}

During normal operation these clocks generate a stream of coded signals that describe various low-level hardware changes relevant for monitoring and steering the accelerator duty cycle (as illustrated in Figure~\ref{fig:numi-time-signals}). For the purposes of this work, these signals are assumed to occur periodically with constant and well-defined frequencies.

\begin{figure}
    \centering
    \includegraphics[width=\linewidth]{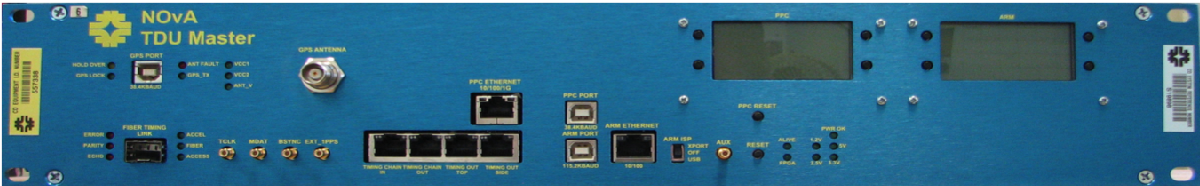}
    \caption{\NOvA{} Timing Distribution Unit (TDU). \cite{norman2012nova}}
    \label{fig:nova-tdu}
\end{figure}

To receive and decode BSYNC and TCLK signals in real-time, \chips{} relies on existing infrastructure developed by \NOvA{}; in particular, its proprietary Timing Distribution System (TDS)~\cite{norman2012nova}. The TDS timestamps accelerator signals with high-precision Coordinated Universal Time (UTC) provided by a commercial GPS satellite receiver, and transmits them through a hierarchy of Timing Distribution Units (TDUs, shown in Figure~\ref{fig:nova-tdu}) that permit system-wide synchronisation to within~\SI{7.8}{\nano\second} across all elements. \chips{} utilised a prototype TDU supplied by \NOvA{}, which was customised for the purposes of this work. Specifically, its on-board PowerPC~8347 computer was reprogrammed to only receive adjustable subset of accelerator signals, convert them from \NOvA{} time specification to International Atomic Time (TAI) and relay them to subsequent stages of the trigger. To allow remote-controlled operation, an independent watchdog system was installed that generates telemetry for \chipsfive{} detector operators, and can reset the TDU upon request.

Following analysis of the accelerator hardware, two signals were selected for consideration with respect to triggering: (1)~TCLK \$A5, which marks the reset of the accelerator prior to the start of a new spill cycle, and (2)~BSYNC MIB\$74, which is generated when neutrinos exit the accelerator complex. These candidates were selected according to the following criteria:%
\begin{enumerate}
    \item
    The selected signal must be reliably generated for all neutrino spills, and not be present otherwise.
    
    \item
    The time elapsed between the selected signal and the neutrino spill must have minimal jitter.
    
    \item
    The time elapsed between the selected signal and the neutrino spill must be sufficiently large, so as to allow \chips{}~DAQ~infrastructure to prepare for the arrival of neutrinos.
\end{enumerate}

%%%%%%%%%%%%%%%%%%%%%%%%%%%%%%%%%%%%%%%%%%%%%%%%%%%%%%%%%%%%%%%%%%%
\subsection{Timing Distribution System}%
\label{sec:signal-delivery-chain}

Once the TDU detects a selected accelerator signal, low-latency notifications are transmitted across the baseline of~\SI{707}{\kilo\meter} to the \chipsfive{} site. This is facilitated by the next stage of the trigger, the \chips{}~Timing Distribution System (TDS). Unlike its \NOvA{}~counterpart, the \chips{}~TDS does not rely on any proprietary components or dedicated network lines. Instead, the system is implemented using conventional computers and widely available network elements, which allow a multitude of relay nodes to communicate using direct socket connections. Sparse spill notifications are multiplexed and transmitted redundantly through these nodes and over the Internet until they reach the \chipsfive{} site, where they are finally processed by the DAQ software that performs sequencing and deduplication.

\begin{figure}
    \centering
    \includegraphics[width=0.8\linewidth]{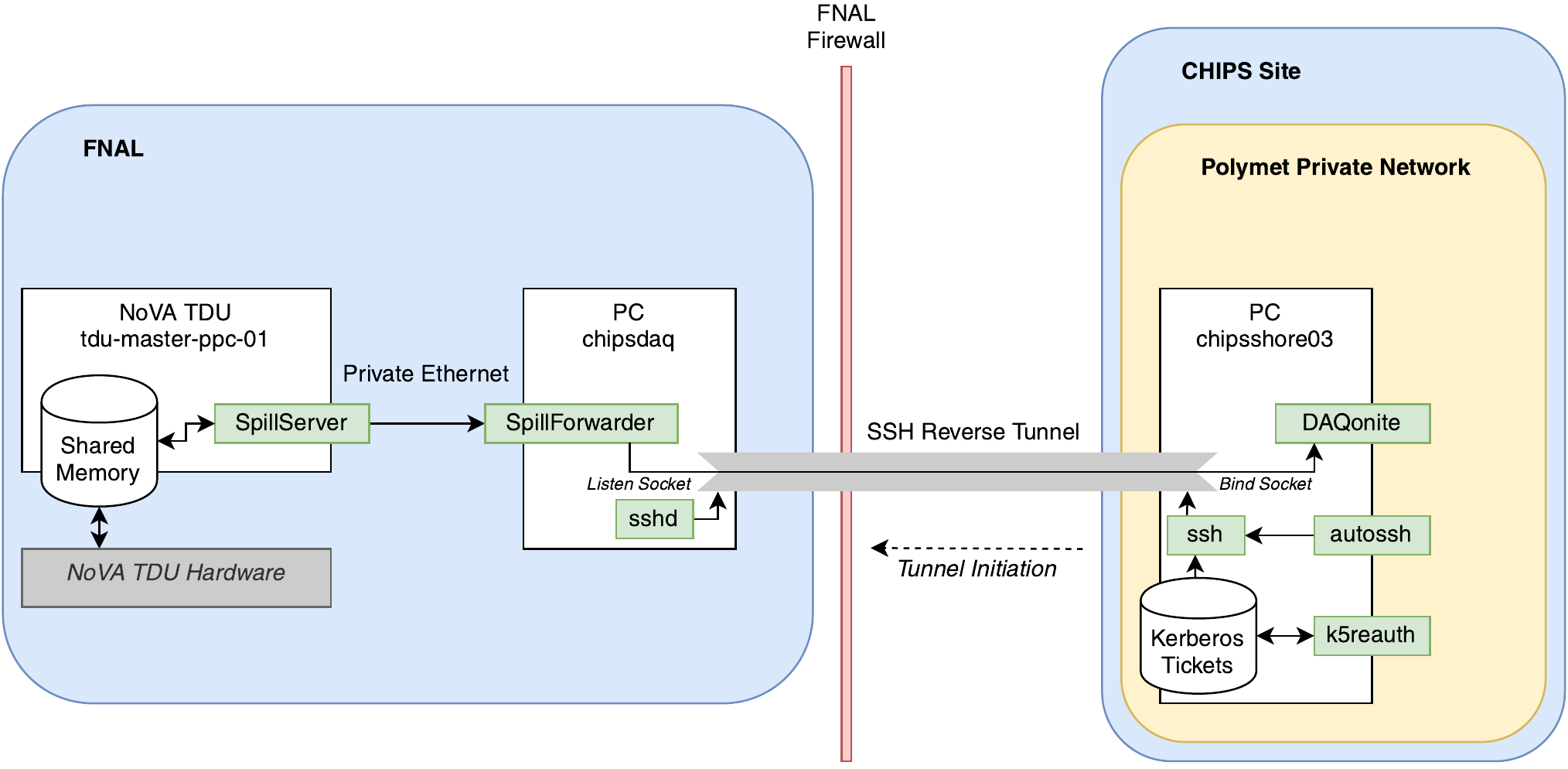}
    \caption{Schematic of the spill signal delivery system. Fermilab accelerator
    signals are decoded by a \NOvA{} TDU, which forwards them through the
    ``chipsdaq'' relay computer to \chips{} detector site.}
    \label{fig:spill-tunnel}
\end{figure}

The \chips{}~TDS must overcome the large physical distance between sites as well as a variety of location-specific network obstacles while maintaining relatively short delivery times. To this end, parts of the system were optimised. The relay node topology installed at Fermilab implements an elaborate tunnelling scheme (illustrated in Figure~\ref{fig:spill-tunnel}) to safely and efficiently traverse the perimeter firewall. At the \chipsfive{} detector site, a reverse Secure Shell (SSH) proxy tunnel was initiated that permits the DAQ~system to expose a direct socket interface reachable from the relay nodes at the Fermilab site. The TDS has been shown to achieve end-to-end latency on the order of \SI{10}{\milli\second}.

%%%%%%%%%%%%%%%%%%%%%%%%%%%%%%%%%%%%%%%%%%%%%%%%%%%%%%%%%%%%%%%%%%%
\subsection{On-site Trigger}%
\label{sec:onsite-triggering}

After spill notifications are delivered to the \chipsfive{} site in Minnesota, they are processed by the last stage of the trigger system which is responsible for integrating spill information with DAQ software in order to promptly apply trigger logic. Since this task is closely coupled with the detector equipment, in addition to trigger integration this section also provides background on the relevant \chipsfive{} DAQ components.

During runs, the detector relies on Plane Optical Modules (POMs) to measure Cherenkov light within the detector volume using photomultiplier tubes (PMTs). POMs digitise analog PMT signals as discrete hits and timestamp them with nanosecond-precise TAI that is synchronised to a local GPS receiver using a hierarchy of White Rabbit devices~\cite{serrano2013white}. Once timestamped, hits are streamed to a data processing facility via a 10~Gbit Ethernet link. Depending on the POM~hardware in use, the presented trigger integrates with this process in two modes:%
\begin{enumerate}
    \item 
    \textit{Software data selection} -- a procedure that performs continuous acquisition but later excludes data taken outside of specified time windows.
    
    \item
    \textit{Hardware triggering} -- a procedure that leaves POMs primed in insensitive standby mode, occasionally instructing them at short notice to become sensitive in order to measure within specified time windows.
\end{enumerate}

\begin{figure*}
    \centering
    \includegraphics[width=0.6\linewidth]{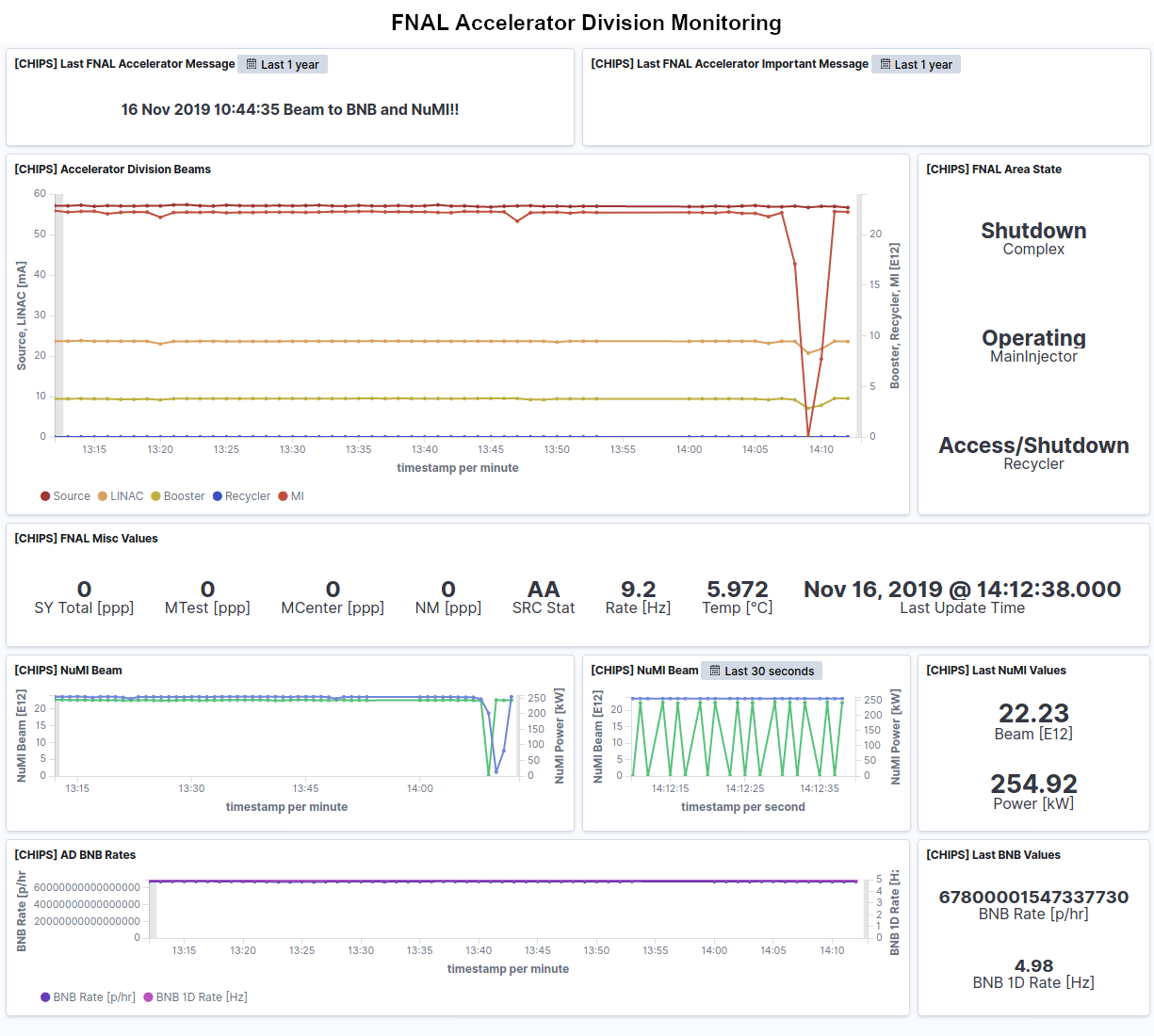}
    \caption{Real-time dashboard that was developed to display
    information about the state of the Fermilab accelerator to \chips{} DAQ
    operators. Visualisation is performed by the Kibana front-end.}
    \label{fig:spill-dashboard}
\end{figure*}

While the former approach requires no hardware upgrades, and is therefore readily compatible with all POMs, the latter approach uses the MicroDAQ subsystem~\cite{huber2018icetop} on-board the latest-generation POMs. Hardware triggering with MicroDAQs is desirable since it saves data bandwidth and reduces DAQ system load, which would otherwise be required to identify and discard out-of-window hits. In both cases, DAQ~software needs to be supplied with a series of triggered time windows in real-time. Depending on run type, these windows are calculated to either include or exclude the estimated time at which the neutrino beam passes through the detector. In parallel, trigger information is fed to on-site monitoring systems that inform detector operators (as illustrated in Figure~\ref{fig:spill-dashboard}).

\section{Benchmark}%
\label{sec:benchmark}
To demonstrate the viability of the presented trigger system, this section evaluates its capability to reliably deliver notifications to the \chipsfive{} detector ahead of the neutrino spills. In this context, it is assumed that each notification and its corresponding spill undergoes the following sequence of events:%
\begin{enumerate}
    \item The trigger signal is detected at the accelerator and a notification is dispatched.
    \item A neutrino spill exits the accelerator complex.
    \item The notification is read at the detector site.
    \item The neutrino spill arrives at the detector.
\end{enumerate}

For convenience, times elapsed between these events are denoted $t_{xy}$, where $x$ and $y$ refer to event numbers~1-4. It is desirable to show that the time elapsed between the arrival of notifications and their corresponding neutrino spills, later labeled as time budget $t_{34}$, is as large as possible, and not smaller than \SI{200}{\milli\second}, which is the estimated minimum time required by DAQ systems to schedule a sensitive time window in hardware triggering mode. To this end, three key properties are investigated:%
\begin{enumerate}
    \item 
    The time difference $t_{12}$ and jitter $\sigma_{12}$ between trigger signal and neutrino spill at the accelerator site.
    
    \item
    The notification latency $t_{13}$ of the \chips{} TDS.
    
    \item
    The time $t_{14}$ between the trigger signal and the arrival of neutrinos at the \chipsfive{} site.
\end{enumerate}

\begin{figure}
    \centering
    {\footnotesize
    \def\svgwidth{0.9\columnwidth}
    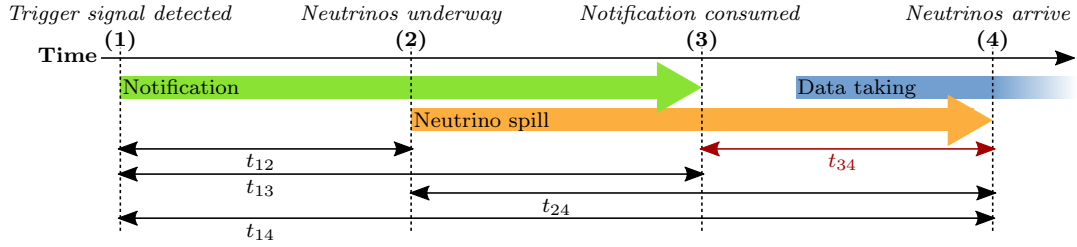
    }
    \caption{Events of interest used in the analysis of a single spill, plotted in the order of time (not to scale). Ongoing processes are indicated using thick colored arrows, whereas time intervals $t_{xy}$ are labeled using thin arrows in the bottom. The time budget $t_{34}$ is indicated in red.}
    \label{fig:event-order}
\end{figure}

These values were chosen to allow the budget~$t_\text{34}$ to be expressed as~$t_{14}-t_{13}$, where $t_{14}$ is substituted for $t_{12}+t_{24}$ (this is also shown in Figure~\ref{fig:event-order}). While~$t_{12},t_{13}$ are measured experimentally, the neutrino time of flight~$t_{24}$ is calculated analytically.

\begin{figure}
    \centering
    \begin{subfigure}[b]{0.6\linewidth}
        \centering
        \includegraphics[height=5cm]{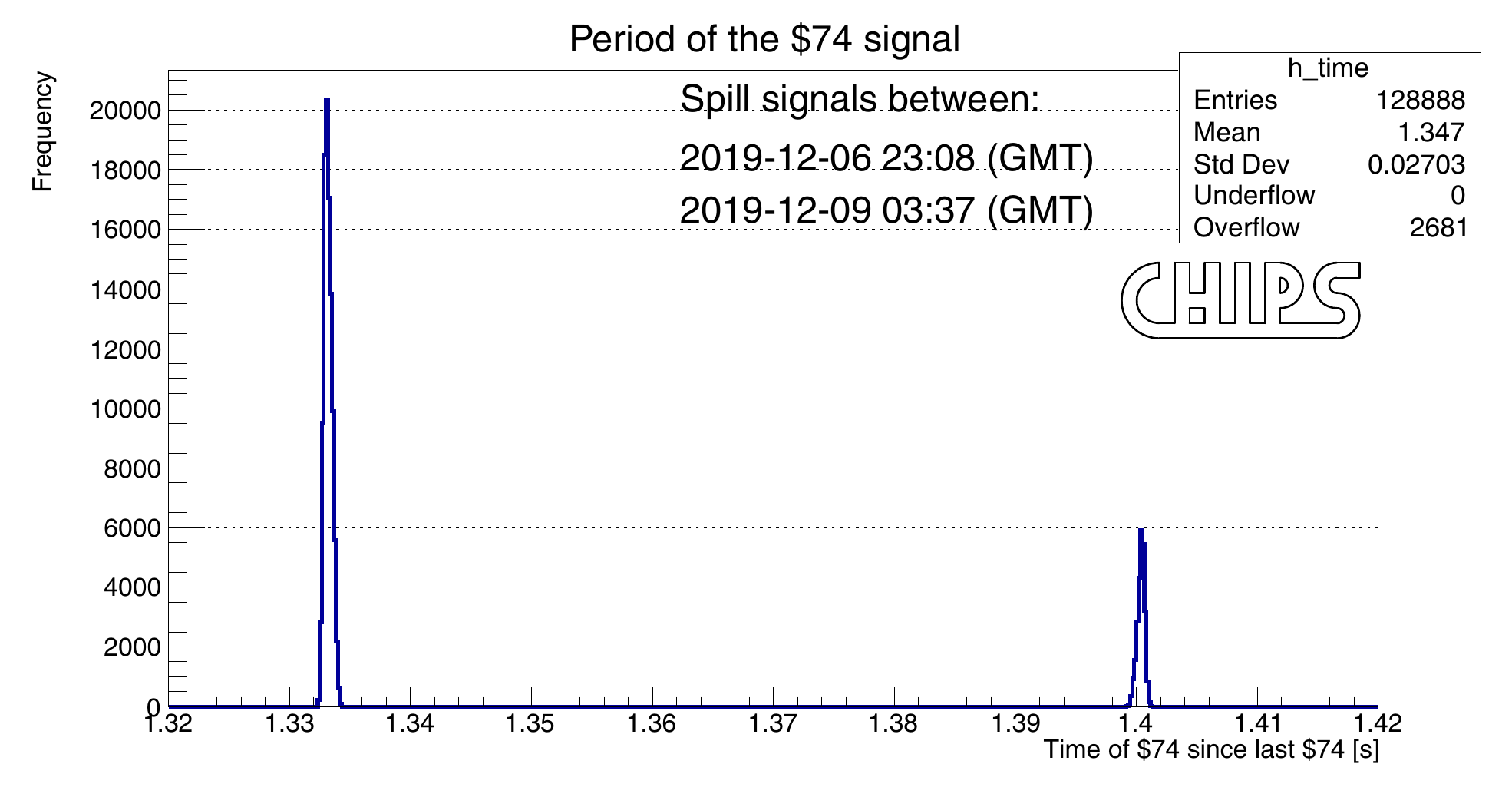}
        \caption{Discrete variations observed in the \$74 signal.}
    \end{subfigure}%
    \begin{subfigure}[b]{0.4\linewidth}
        \centering
        \includegraphics[height=5cm]{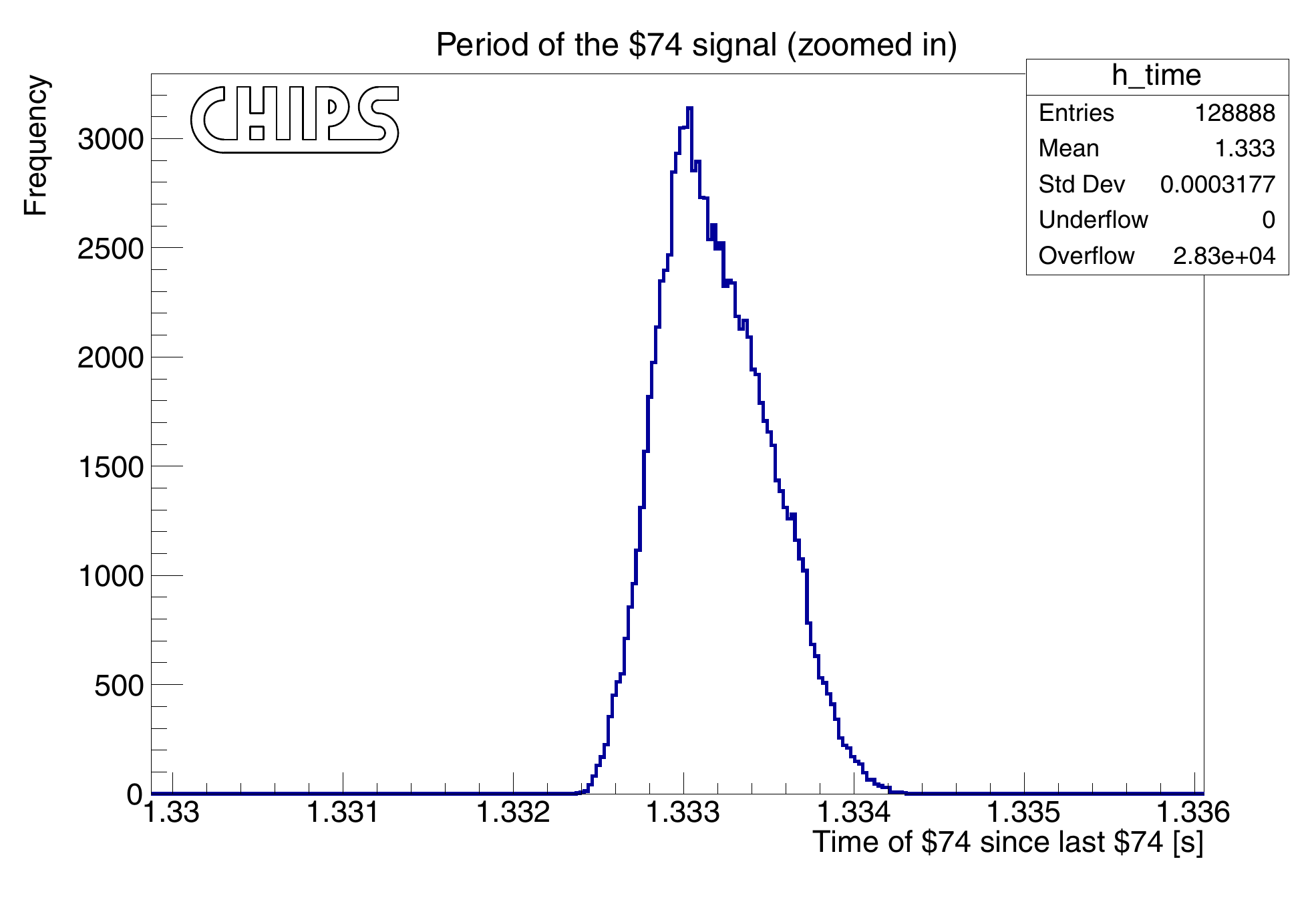}
        \caption{Detail of the peak at \SI{1.333}{\second} in plot (a).}
    \end{subfigure}
    \caption{Time between subsequent emissions of the \$74~signal aggregated in a histogram over the period of roughly 52~hours of
    NuMI beam operation.}
    \label{fig:s74-since-s74}
\end{figure}

Before proceeding to evaluation, it is first necessary to verify the instrumentation used for this analysis; specifically, the customised \NOvA{} TDU (introduced in section~\ref{sec:nova-tdu}). This is achieved by measuring the period of the BSYNC MIB\$74 signal, and comparing it to the expected value\footnote{Even though the observed period is known to vary in discrete increments depending on the programmed accelerator cycle, the given value can be expected to be dominant.} of \SI{1.333}{\second}. Aggregating roughly \num{128000}~real signals observed over a 3-day run during NuMI operation in late~2019 (shown in Figure~\ref{fig:s74-since-s74}), the period was experimentally determined to be $\SI{1.333}{\second}\pm\SI{317.7}{\micro\second}$. This agreement was found to be acceptable for the purposes of this study.

While \$74 would represent an ideal choice for a trigger signal since it marks the precise time of neutrino exit from the accelerator complex, it is considered unsuitable because it is emitted late in the duty cycle, and cannot be delivered to the detector with a sufficient time budget. For this reason, the rest of the analysis assumes the trigger signal (event~1) to be TCLK \$A5, which appears earlier. In addition, \$74~is understood to be the point of departure for the neutrino time-of-flight calculation (event~2). With this interpretation established, we may now proceed to estimate the three key properties of interest introduced at the beginning.

\begin{figure}
    \centering
    \begin{minipage}{0.48\textwidth}
        \centering
        \includegraphics[height=5.4cm]{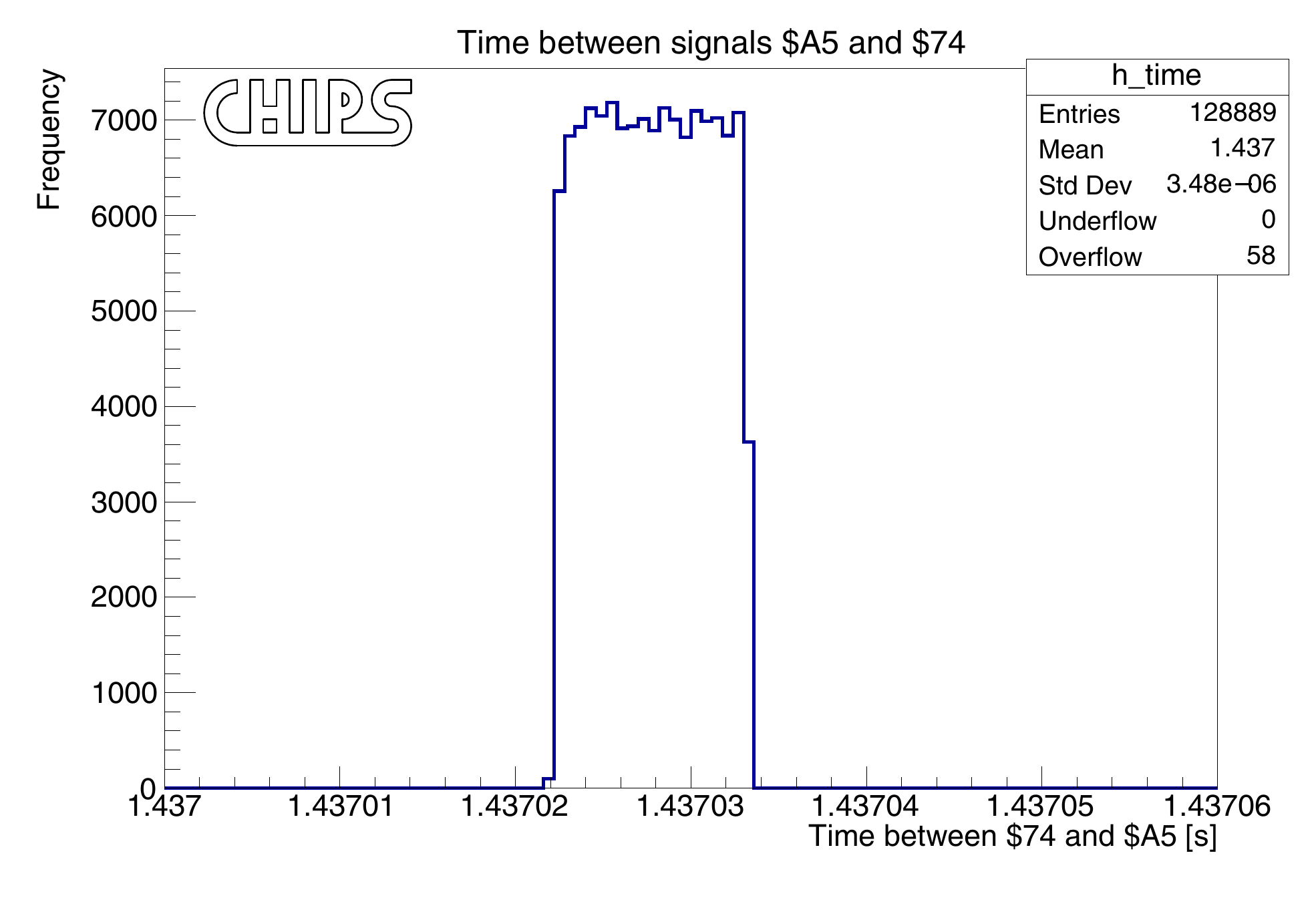}
        \caption{Time~$t_{12}$ between emission of an~\$A5 signal and subsequent~\$74 signal
            aggregated in a histogram over the period of roughly 52~hours of NuMI beam operation.}
        \label{fig:s74-since-sA5}
    \end{minipage}\quad
    \begin{minipage}{0.48\textwidth}
        \centering
        \includegraphics[height=5.4cm]{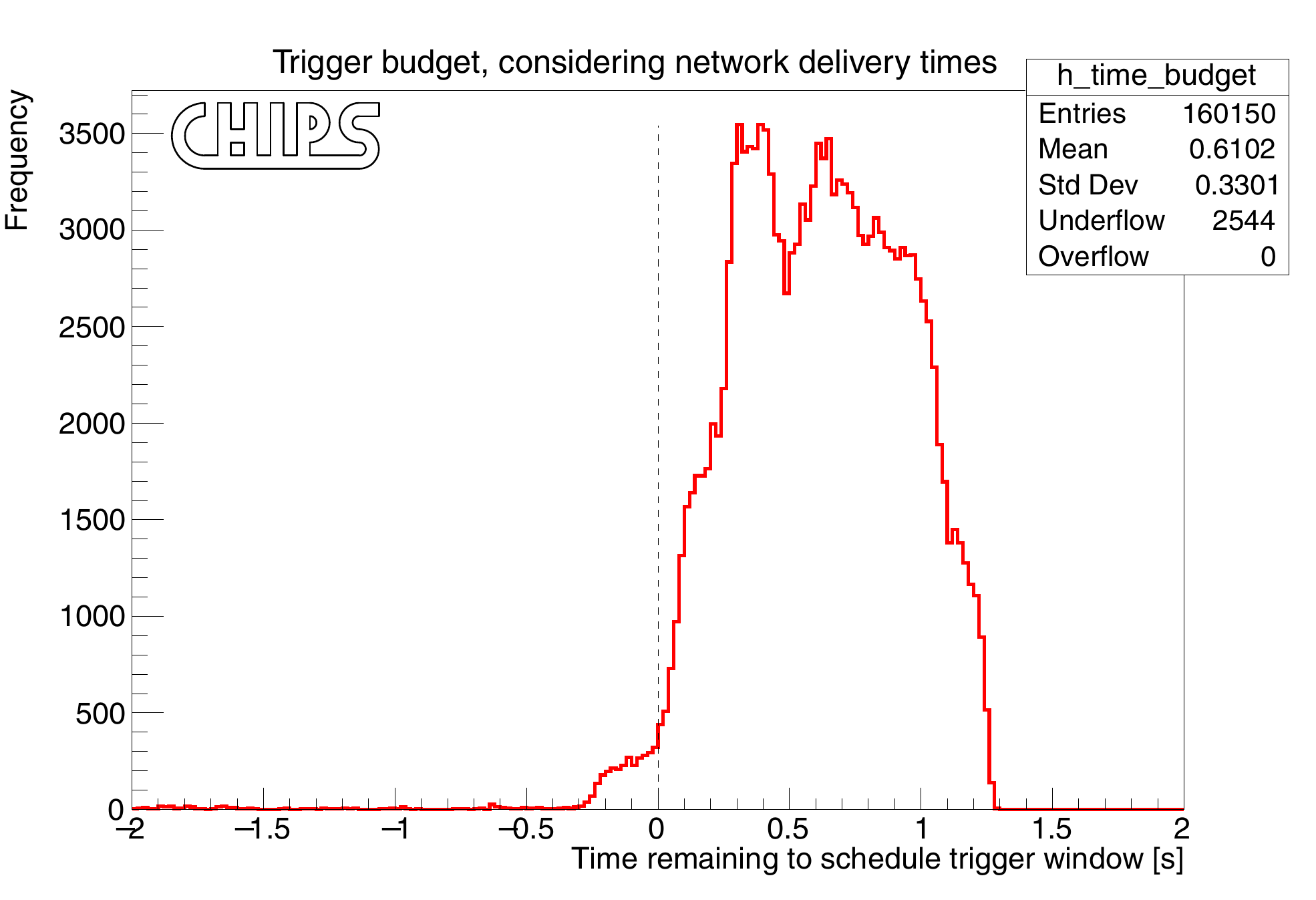}
        \caption{Histogram of the budget~$t_{34}$ remaining to schedule time window in \chips{} DAQ. Negative values do not allow triggering since neutrinos arrive at the detector prior to notification.}
        \label{fig:spill-time-budget}
    \end{minipage}
\end{figure}

The time $t_{12}$ is measured by examining subsequent occurrences of \$A5 and \$74 (as shown in Figure~\ref{fig:s74-since-sA5}). Processing the same data that was previously used for instrument verification, $t_{12}$ was found to be \SI{1.437}{\second} with jitter $\sigma_{12}=\SI{3.48}{\micro\second}$. In order to measure~$t_{13}$, the \chips{}~TDS was modified to assign two additional timestamps to each spill notification, one at the time of its transmission, and another at the time of its consumption by DAQ. The value of~$t_\text{13}$ is obtained as the difference between these timestamps. Finally, $t_{24}$~is analytically given by~$d/c$, where $d=\SI{707}{\kilo\meter}$ is the baseline length and $c$ is the speed of light.

% In order to evaluate $t_\text{budget}$, two more components remain: notification latency~$t_\text{notify}$ and neutrino time of flight $t_\text{ToF}$, which together with~$t_{\$A5\rightarrow\$74}$ contributes to~$t_\nu$. 

% In order to assess trigger signal viability, the amount of time from its emission to the subsequent arrival of neutrinos at the \chipsfive{} detector must be considered. This period represents the \textit{total time budget}, during which the signal must be delivered between sites in order to remain viable for triggering. The budget is constituted by two components: (1) the time from the signal emission to the neutrino spill at the accelerator site $t_{\$A5\rightarrow\text{spill}}$, and (2) the neutrino time of flight $t_\text{travel}$ that is given by~$d/c$, where $d=\SI{707}{\kilo\meter}$ is the baseline length and $c$ is the speed of light. Since the \$74 signal marks the moment of the accelerator neutrino spill, the duration between two subsequent signals $t_{\$A5\rightarrow\$74}$ known from the accelerator duty cycle can be used to calculate $t_{\$A5\rightarrow\text{spill}}$ in theory.

Combining all components together, the time budget $t_{34}=t_{14}-t_{13}$ may now be evaluated. As mentioned earlier, the time~$t_{14}$ is substituted for $t_{12}+t_{24}$. To obtain a conservative lower bound on $t_{14}$, the measured jitter $\sigma_{12}$ is subtracted from $t_{12}$, yielding the following result:%
\begin{align}
    t_{14}&=t_{12}+t_{24} \\
          &\geq t_{12} - \sigma_{12} + t_{24}\\
                   &\approx\SI{1.437}{\second} - \SI{3.48}{\micro\second} +
                   \SI{2.5}{\milli\second}
                   =\SI{1.4395}{\second}.
\end{align}

% Alternatively, the same duration can also be measured experimentally (as shown in Figure~\ref{fig:s74-since-sA5}) with the added benefit of considering the jitter $\sigma_{\$A5\rightarrow\$74}$ in the calculation. In such case, the estimate is conservatively given by~$t_{\$A5\rightarrow\$74}-\sigma_{\$A5\rightarrow\$74}$. Combining all the listed components, the time budget evaluates as

% After subtraction of the signal delivery time due to network latency, the remaining budget for scheduling time windows at the \chips{} detector site based on this calculation is plotted in Figure~\ref{fig:spill-time-budget}.

With this value, the time budget $t_{34}$ experimentally evaluates as $610.2\pm\SI{330.1}{\milli\second}$ (plotted in Figure~\ref{fig:spill-time-budget}). Out of roughly \num{160000} observed signals, approximately \num{3.69}~\% had negative time budget, meaning that in such cases notifications were delivered to the detector only \textit{after} it had already encountered NuMI neutrinos. The remaining \num{96.3}~\% notifications preceded their corresponding neutrino spills. Furthermore, $t_{34}$ as well as $t_{34}-\sigma_{34}$ exceed the required \SI{200}{\milli\second}~threshold imposed by DAQ hardware, meaning that more than \num{83.9}~\% of all notifications will be viable.

\section{Conclusion}%
\label{sec:conclusion}

A new triggering system was developed that allows delivery of nanosecond-precise accelerator signals from Fermilab site to the \chipsfive{} detector \SI{707}{\kilo\metre}~away. Aside from existing \NOvA{} infrastructure, the presented system relies only on open-source software, conventional computing and network infrastructure to provide a resilient low-latency trigger.

In a benchmark, the system delivered over 96~\% notifications ahead of their corresponding neutrino spills, and more than 83~\% notifications with a sufficient time budget required for hardware-assisted triggering. This permits the latest-generation \chips{} POMs to save data bandwidth and reduce DAQ computing load outside of triggered time windows.

\section*{Acknowledgements}

This work was supported by Fermilab; the Leverhulme Trust Research Project Grant; U.S.~Department of Energy; and the European Research Council funding for the CHROMIUM project.
Fermilab is operated by Fermi Research Alliance, LLC, under Contract No. DE-AC02-07CH11359 with the U.S.~DOE.

\bibliographystyle{elsarticle-num}
\bibliography{references.bib}

\end{document}